# Evaluation of blood coagulation by optical vortex tracking

Yaowen Zhang, [†a] Jiaxing Gong, [†a,b] Hui Zhang, [a] Qi Li, [a,b] Guangbin Ren, [a] Wenjian Lu, [a] and Jing Wang[a,b,*]

[a]Huazhong University of Science and Technology, College of Life Science and Technology, Biomedical Engineering Department, Wuhan, China
[b]Shenzhen Huazhong University of Science and Technology Research Institute, Shenzhen, China

**Abstract**
We investigated a new optical approach for the detection of the coagulation dynamic process by means of the information on the optical vortex. In our study, laser speckle was captured using a high-speed CMOS camera, and the statistical information of the optical vortex characterized the change in coagulation properties with time. Similar to the scattering particles, the motion of the optical vortex is restricted during coagulation, as a result, the whole process of coagulation can be detected by calculating the mean square displacement(MSD) of the optical vortex. The results demonstrate a close correlation between coagulation parameters measured using the optical vortex method and thrombelastography(TEG), creating a powerful opportunity for self-testing and real-time detection of coagulation.

[†]These authors contributed equally to this work.
*Corresponding author**, E-mail: wang.jing@hust.edu.cn

## 1  Introduction

The coagulation system is the body's defense mechanism for maintaining normal blood flow and preventing blood loss.[1] Its abnormality can result from a variety of conditions including severe trauma, temperature imbalance, or surgery, and can cause coagulopathy including subcutaneous bleeding, atherosclerosis, and even be life-threatening.[2-4] The normal coagulation process depends on a delicate dynamic balance among the coagulation, anticoagulation, and fibrinolysis. Therefore, it is important for diagnosis, prevention, and treatment of diseases to monitor the coagulation status dynamically.

However, coagulation is a cascade of complex biochemical processes due to the interactions between platelets, red blood cells, and more than a dozen coagulation factors, as well as two different coagulation pathways, endogenous and exogenous.[5-7] Conventional coagulation tests(CCTs) generally assess a patient's coagulation status by measuring the activated thromboplastin time (aPTT), prothrombin time (PT), platelet count (PLT), and fibrinogen levels, which requires a lot of time to separate plasma, platelets and other components from the patient's blood. Thus, traditional tests are not efficient enough and the results are only the coagulation parameters at a certain moment in time, which can't reflect the dynamics process of coagulation. During surgery, trauma bleeding casualties remain the leading cause of death.[8] Therefore, novel methods that enable the assessment of the coagulation dynamics process efficiently and promptly become essential.

The main clinical devices that can monitor the coagulation dynamics process are thrombelastography (TEG) and rotational thrombelastography (ROTEM) currently. They can assess the formation, dissolution process, and strength of the blood clot by applying a continuous oscillation to the blood sample and contacting it with a metal wire.[9,10] While they overcome the drawback of CCTs such as inefficient and inability to detect the whole process of coagulation, its methods based on contact with blood and application of excitation can delay clot formation and modify clot structure during measurements, thereby providing inaccurate results.[11] Moreover, the sophisticated components and complex structure of TEG and ROTEM equipment make it less operable, more difficult to maintain, and more expensive.

To avoid direct contact with the sample, microrheological methods are also often used to examine the rheological properties of soft tissues. According to the Generalized Stokes-Einstein Relation(GSER), the viscoelasticity of a complex fluid can be obtained by calculating the MSD of the probe particles in it. Based on microrheology, laser speckle rheology(LSR) can calculate the viscoelasticity of the environment that determines the motion of the scattering particles by calculating the variation of the laser speckle with time formed by multiple scattering, which reflects information on the motion of the scattering particles.[12-16] Thus, this method has been applied to the measurement and characterization of viscoelasticity in biological tissues such as blood, articular cartilage, mammary glands, etc. When coherent light is irradiated on a multiple scattering medium of biological tissues such as whole blood, coherent light scattered by the scattering particles in the medium will form a spatially distributed pattern of bright and dark granular spots, i.e. laser speckle, as shown in Fig. 1(a). The temporal evolution of the laser speckle contains the information of scattering particles' motion. The laser speckle change with the movement of scattering particles and the faster the particles move, the faster the laser speckle changes. Furthermore, the Brownian motion of scattering particles is directly related to the viscoelastic properties of the medium. Thus, LSR can characterize the viscoelasticity of biological tissues by measuring the time scale of speckle intensity fluctuations, $g_2(t)$. However, the measurement of $g_2(t)$ in practice is very complicated and requires consideration of static scattering, ordered or disordered motion of the scattering particles, the number of times the coherent light is scattered, the offset caused by noise, etc,[17] which is not equivalent to the result simply by a single model or formula as in LSR theory.

The laser speckle has a unique structure known as the optical vortex, which exists in the intensity of zero.[18] The real and imaginary parts of the amplitude of the optical vortex are thereby both zero, so the optical vortex is also named "phase singularities". The phase around the optical vortex increases or decreases around the central spiral. According to Freund's sign rule of the optical vortex, the optical vortex whose phase increases counterclockwise is called "positive", and the optical vortex whose phase increases clockwise is called "negative".[19] Moreover, the positive and negative optical vortex always appear and disappear in pairs. As shown in Fig. 1(b), the red point is the positive optical vortex and the green point is the negative optical vortex. The optical vortex is an important research topic in the field of optics and is gradually forming a branch of exotic optics, which has significant applications in many optical fields such as optical measurement, micro displacement measurement, optical communication, and optical imaging.[20-28]

Based on the previous research, we find that the motion of the optical vortex directly reflects the motion of the scattering particles that are related to the viscoelasticity of the medium. The statistical properties of the optical vortex's motion are thereby able to characterize the viscoelasticity of the medium. In this study, we propose a novel method for the evaluation of blood coagulation dynamics. We first performed coagulation experiments on porcine blood, calculated the MSD of the optical vortex during coagulation, and investigated the relationship between the change in the MSD of the optical vortex and blood coagulation dynamics. Subsequently, the coagulation parameters including reaction time(R), activated clotting time(ACT), maximum viscoelastic modulus of the thrombus, which can reflect the dynamic process of coagulation, were obtained by feature extraction of coagulation curves. The accuracy of these parameters in reflecting the dynamic process of coagulation was verified by correlation analysis of these parameters with corresponding clinical indicators, such as those measured by thromboelastography. Finally, two-dimensional images of the viscoelasticity of blood were obtained by two-dimensional scanning during coagulation, which is potential to be a new approach to research the spatial variation of blood viscoelasticity during coagulation.

## 2 Materials and methods

*2.1 Porcine blood specimens and coagulation assay preparation*

We examined the coagulation dynamics process of two types of porcine blood anticoagulated with sodium heparin and sodium citrate by the OVM and compared it with TEG. Different coagulation parameters were obtained by varying the amount of blood used and the amount of calcium chloride. Briefly, each blood sample was warmed to 37 °C in a water bath for 5 minutes and 800~1200 µL of blood was pipetted into vials pre-loaded with a kaolin buffer solution. After 3 gentle vial inversions, 0.2M calcium chloride(35~70 µL) was added to the kaolin vial for activating coagulation. Continue 3 gentle vial inversions, ~120 µL of the kaolin-activated whole blood was immediately loaded into an imaging chamber(disposable imaging cartridge ), in the meantime, an additional 360 µL blood from the same tube was loaded into the TEG cup. To facilitate an accurate comparison between the OVM and TEG, measurements were conducted simultaneously.

*2.2 Experimental setup*

The optical experimental setup in Fig. 1(C) was used to capture time-varying laser speckle patterns of the prepared dynamic specimens. Light from a laser diode source was reflected off a beam splitter and focused on the top surface of the blood sample in an imaging chamber with an objective lens. The heat plate maintained the temperature at 37 °C during the process of blood coagulation. Laser speckle patterns were formed by the interference of the backscattered light passing through the medium and were acquired by a high-speed CMOS camera, together with a 4-f system and a linear polarizer. The speckle size is approximately 4 × 4 pixels to satisfy the Shannon sampling theorem.[29] The beam dump was used to contain the unwanted laser beam. For each sample, measurements were performed for a total imaging duration of 20 minutes to evaluate the coagulation dynamics process. At every 20 seconds, one speckle frame sequence was captured at a frame rate of 800 frames per second(RoI: 192 × 192 pixels) for 0.5 seconds. Thus, a total of 60 time series of 400 speckle frames were collected over 20 minutes of the experiment.

To demonstrate the capability of the OVM for monitoring micro thrombosis, the Spatio-temporal analysis was used with the motion controller. The optical setup was reconfigured for satisfying the experiment requirements.

*2.3 Calculating the MSD of the optical vortex during blood coagulation*

In the previous research, We found that the motion of the optical vortex is directly related to the Brownian motion of scattering particles in a multiple scattering medium, and the statistical properties of the optical vortex motion can serve as indicators for the viscoelastic properties of the medium. Here we utilized the MSD of the optical vortex to estimate the evaluation of blood coagulation dynamics:

$$\langle \Delta r^2(t) \rangle = \langle |\vec{r}(t_0 + t) - \vec{r}(t_0)|^2 \rangle \quad (1)$$

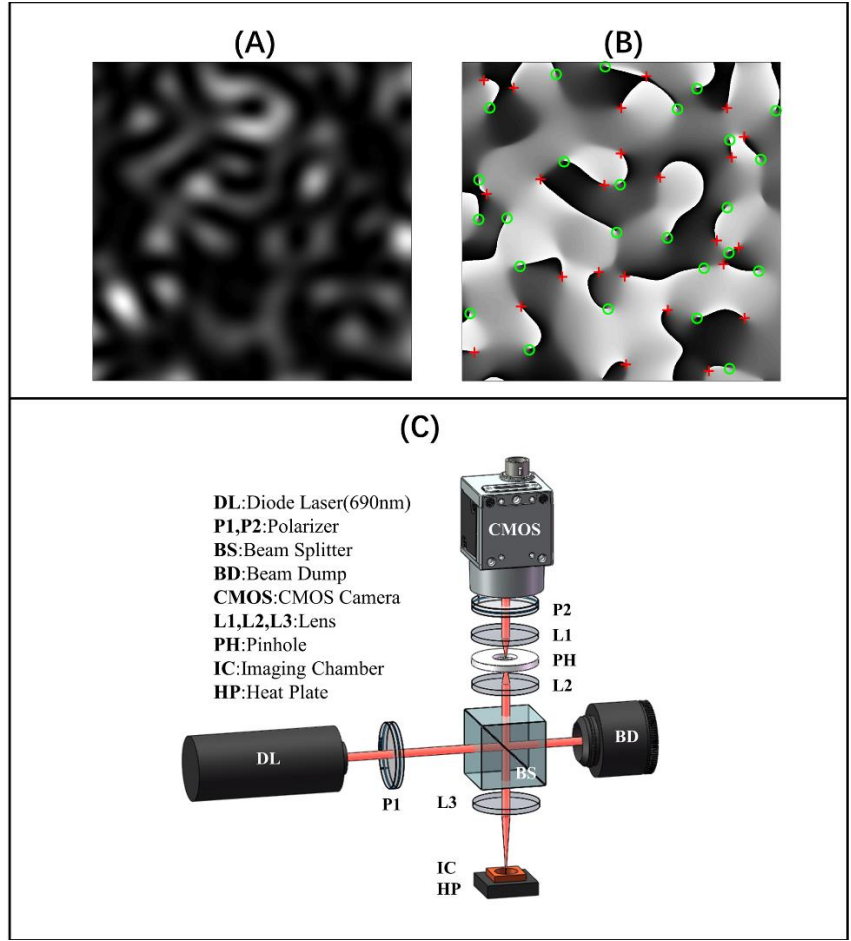

Fig. 1. (A) Laser speckle pattern captured from porcine blood. (B) Pseudo-phase of the corresponding (A). The red points are positive optical vortices and the green points are negative optical vortices. (C) The instrument of the optical vortex setup. Light from a laser diode source(690 nm, 30mW, Newport Corp.) was reflected off a beam splitter and focused on the customized imaging chamber containing ~120 μL of porcine blood with an objective lens(5X). In the back-scattering geometry, laser speckle patterns were acquired by a high-speed CMOS camera (acA2000-340kmNIR, Basler AG), together with a 4-f system and a linear polarizer.

Where $\vec{r}(t_0 + t)$ and $\vec{r}(t_0)$ represents the positions of the optical vortex at times $t_0 + t$ and $t_0$, and $<>$ indicates ensemble averaging over the recording duration of a single speckle sequence. For locating the optical vortex while reducing the complexity of the optical system, the captured speckle sequences were first converted into the complex analytic signal by Laguerre-Gaussian transform to obtain pseudo-phase $\hat{\phi}$.[21] The formula of transform as follows:

$$\tilde{I}(x,y) = |\tilde{I}(x,y)| \exp[i\hat{\phi}(x,y)] = I(x,y) * LG(x,y) \quad (2)$$

Where $I(x,y)$ is the original speckle intensity distribution and $\tilde{I}(x,y)$ represents the corresponding complex analytic signal. $\hat{\phi}(x,y)$ is the pseudo-phase representation, $LG(x,y)$ represents the Laguerre-Gaussian function in the spatial signal domain, which can be calculated by the inverse Fourier transform of formula (3):

$$LG(f_x, f_y) = (f_x + if_y)\exp\left(-\frac{f_x^2 + f_y^2}{\omega^2}\right) = \rho \exp\left(-\frac{\rho^2}{\omega^2}\right)\exp(j\beta) \tag{3}$$

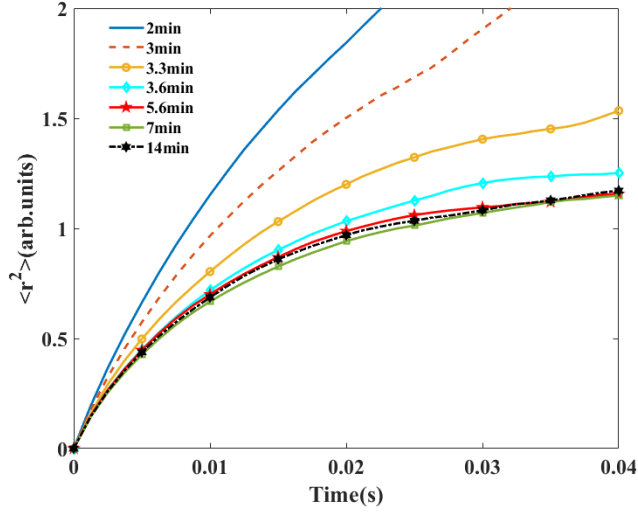

Fig. 2 MSD curves of optical vortex calculated at 2, 3, 3.3, 3.6, 5.6, 7, 14 minutes during blood coagulation. It was obvious that the MSD curves decreased sequentially as the coagulation time proceeds. And the rate of decay slowed down at the same time. A complete plateau level was reached after coagulation.

$\rho = \sqrt{f_x^2 + f_y^2}$ and $\beta = \arctan(f_y/f_x)$ are the polar coordinates in the frequency domain. $\omega$ represents the bandwidth of the Laguerre-Gaussian function, which can adjust the size of the speckle grain.[21, 30] $*$ is the convolution operator. The pseudo-phase $\hat{\phi}(x, y)$ can be Calculated such as:

$$\hat{\phi}(x, y) = tan^{-1}\frac{Im\{\tilde{I}(x, y)\}}{Re\{\tilde{I}(x, y)\}} \tag{4}$$

Next, the optical vortex was located according to equation (5)

$$n_t \equiv \frac{1}{2\pi}\oint_c \nabla\hat{\phi}(x, y) \cdot d\vec{l} \equiv \frac{1}{2\pi}\iint_D \left(\frac{\partial^2\hat{\phi}}{\partial x \partial y} - \frac{\partial^2\hat{\phi}}{\partial y \partial x}\right)\partial x \partial y \tag{5}$$

where $n_t$ is the topological charge of the optical vortex, $\nabla\hat{\phi}(x, y)$ defines the local phase gradient and the contour integral is taken over path l on a closed-loop $c$ around the vortex. It is clear that $n_t=0$ everywhere $\hat{\phi}$ is differentiable, except at the optical vortex locations where the phase is undefined. We used a 2-dimensional phase unwrapping function to determine the position and topological charge of all the optical vortex with a window of $2 \times 2$ pixels at every point of the measured pseudo-phase map, the spatial accuracy limited by the CMOS pixel size. Only the topological charges of $\pm 1$ are recorded. To calculate the MSD of the optical vortex according to formula (1), we identified the corresponding optical vortex between two adjacent phase maps using restrict the search not to exceed the average diameter of speckle grain. MSD curves were calculated for each speckle sequence time series and plotted over the entire blood clot duration time, $0 < t \leq 20$ minutes (Fig. 2).

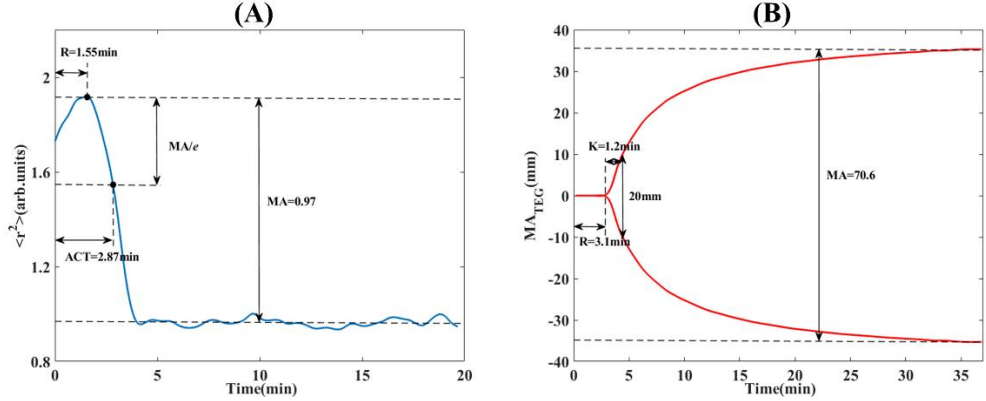

Fig. 3. Coagulation time curves of the same sample measured by the OV method and TEG. (A) Time trace of MSD at the moment of 0.02s. (B) Corresponding coagulation profile measured by TEG. Differences in coagulation parameters result from different methods.

## 2.4 Calculating Coagulation Parameters

As previously described, the MSD of the optical vortex is able to characterize the viscoelasticity of the medium. We measured all the MSD data at the moment of 0.02s to indicate clot viscoelasticity for the whole coagulation process. To calculate the coagulation parameters, the time course of the MSD data was displayed as a function of coagulation time, t. (Fig. 3. (A)). Next, the time trace of blood coagulation dynamics measured using TEG during coagulation was compared with it measured using the OVM. (Fig. 3. (B)). From the clotting curve by the OVM, the following coagulation parameters were extracted for comparison with TEG: reaction time (R), Kinetic time (K), activated clotting time (ACT =R+K), and maximum amplitude (MA). In the OVM, the R time was defined as the first distinct turning point on the clotting curve. The MA represents the maximum stiffness of the clot, as defined by the peak amplitude of the clotting curve. And the ACT time denoted the time to maximum fibrin formation, was measured by calculating $1/e$ of the amplitude. The Angle reacts to the rate of clot polymerization, which has the same meaning as indicated by the K value. Thus, We ignore the calculation of Angle.

In the TEG method,[9] the time between the start of data collection (at the 'SP time') to amplitude greater than 2mm is defined as the reaction time (the 'R time')—that is deemed necessary to initiate the fibrin network formation. The 'K value' is denoted the time taken from the beginning of clot formation until the amplitude of the TEG output trace reaches 20mm. The measurement of MA is similar to the OVM, but the MA of TEG is directly characteristic viscoelasticity of the blood sample.

## 2.5 Statistical Analysis

To validate the sensitivity of the OVM, we conducted 100 experiments by varying the amount of blood used and the amount of calcium chloride. The amount of blood used was 800-1000 µL and the amount of calcium chloride was 30-80 µL. After coagulation experiments with two types of anticoagulated porcine blood with sodium heparin and sodium citrate, we obtained coagulation parameters with a relatively wide distribution of values.

From the 100 different coagulation experiments, the correlation between OV and TEG coagulation parameters, R, ACT, and MA was calculated using the parametric Pearson's correlation coefficient, r, analysis.

## 3  Results

Figure. 2 shows the MSD curves of the optical vortex reported at 2, 3, 3.3, 3.6, 5.6, 7, and 14 minutes following kaolin activation of blood coagulation. It is clear that the changing trend of MSD measured by the OVM is related to the viscoelastic properties of the clotting blood sample. As the coagulation reaction proceeds, the MSD of the optical vortex decreased sequentially. Following re-calcification and kaolin activation of coagulation factors, i.e. the initial stage of coagulation reaction, the MSD curves decayed at a fast rate(2-3.3 minutes). At 3.6 minutes after the onset of coagulation reaction, the MSD tends to level. A complete plateau level was reached at 5.6 minutes, and the subsequent measured MSD curves were maintained in its vicinity. Thus, after the blood coagulation process attained equilibrium characterized by a fully-formed, fibrin-platelet, the MSD curve changed little.

The blue curve in Fig. 3. (A) shows the results of the clotting curve measured by the MSD data of the optical vortex in the porcine blood anticoagulated with sodium citrate during coagulation. The total duration of the OVM measurement is 20 minutes. The corresponding values measured for the same blood sample using TEG have been plotted red in Fig. 3. (B). TEG's trace is plotted by the displacement of an inner cylinder (the 'pin') suspended on a torsion wire. Because the measurement of the MA parameter is slower than the OVM, TEG's measuring time is over 30 minutes. As observed, the trends in the clot viscoelasticity profiles measured by the OVM closely resembles the TEG trace. Before the reaction time of coagulation, MSD exhibited only a little change(From 1.7 at 0 minutes to 1.9 at 1.55 minutes) while the amplitude of TEG maintained

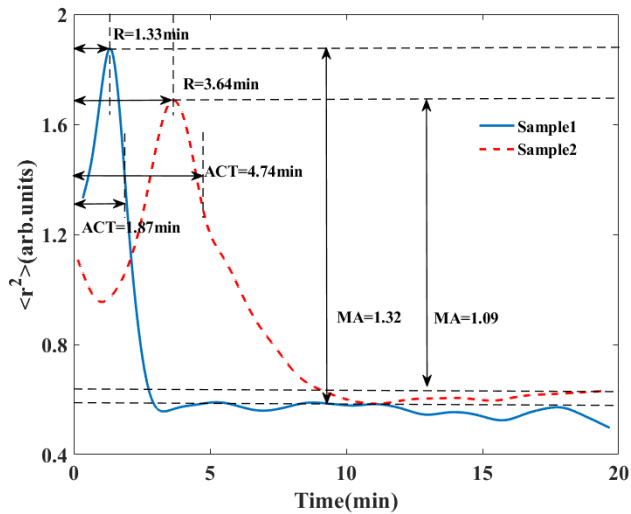

Fig. 4 Time traces of coagulation measured by using 35μL and 70μL 0.2M calcium chloride. It is clear that both R and ACT were significantly prolonged when calcium chloride used was from 70 μL decreased to 35 μL. In the contrast, the MA decreased as the calcium chloride used decreased.

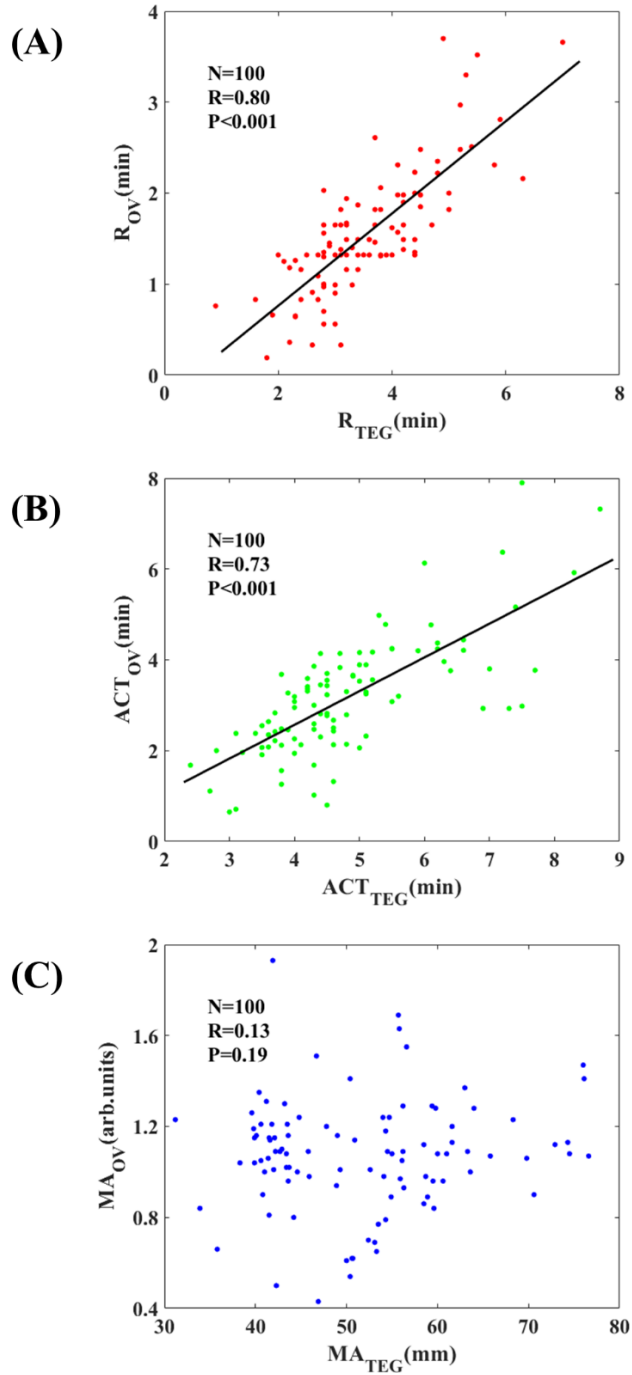

Fig. 5. (A) R-time, (B) ACT-time, and (C) MA measured by the OV method and TEG in 100 different porcine blood experiments. (A) and (B) showed a strong positive correlation with the Gold standard but (C) demonstrated a weak correlation, which is worthy of discussion.

0(From 0 minutes to 3.1 minutes). This indicates that initially, blood exhibits the characteristics of a viscous material of low modulus, which restricted the movement of the optical vortex less. During the progression of coagulation, the value of MSD decayed rapidly from 1.9 at 1.55 minutes to 0.96 at 4.15 minutes. But the amplitude of TEG grew slower from 0 mm at 3.1 minutes to 35.3mm at 36 minutes. This indicates that the viscoelasticity of the blood increased gradually which increasingly restricted the movement of the optical vortex. Following stabilization of the coagulation process at later clotting times, both MSD and TEG amplitude approached saturation levels. It is worth noting that the measured total clotting time and R time by the OVM are less than those by TEG. It is likely due to the higher sensitivity in detecting the optical vortex movements resulting from alterations in clot modulus during coagulation. Furthermore, the TEG's principal drawback is that the strain amplitude associated with its operation is uncontrolled and decreases progressively during coagulation, which can substantially delay clot formation and modify clot structure during measurements.[11]

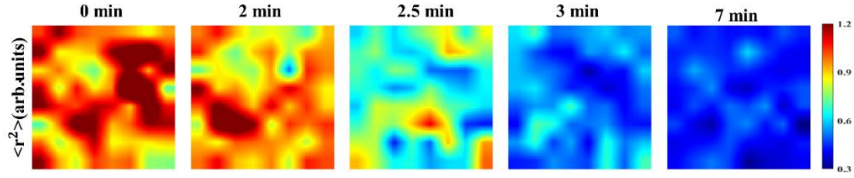

Fig 6 Spatial maps of MSD at minutes 0, 2, 2.5, 3, 7 after kaolin activation. Significant MSD values change appear at 2-2.5 minutes and maintain a low level over 7minutes in the porcine blood.

Fig.4 shows coagulation experiments using 35 µL and 70 µL, 0.2M calcium chloride, with other conditions such as the amount of blood used and types of anticoagulants remaining consistent. As an essential factor in the coagulation reaction, the amount of calcium directly affects the whole coagulation process. Thus, different coagulation parameters were obtained by controlling the amount of calcium used. As shown in the figure, R time is 1.33 minutes and ACT is 1.87 minutes by using 70 µL calcium chloride. When the calcium chloride used was decreased to 35 µL, both R and ACT were significantly prolonged to 3.64 minutes and 4.74 minutes. This indicates that increasing the amount of calcium chloride within a certain range can accelerate the clotting reaction. Meanwhile, the amplitude(MA=1.32) measured by 70 µL calcium chloride used is larger than it(MA=1.09) by 35 µL calcium chloride. It is can be explained that more calcium chloride led to more thrombin which can convert fibrinogen to fibrin, enabling the formation of a fibrin clot.

In Fig. 5, the coagulation parameters, R, ACT, and MA, measured from the OVM of 100 different porcine blood experiments are plotted against the corresponding parameters measured by the TEG device. The coagulation parameters were measured by the OVM with a relatively wide distribution of values such as R ranging from 0.19 to 3. 7 minutes, ACT ranging from 0.65 to 7.9 minutes, and MA ranging from 0.43 to 1.93. And the corresponding coagulation parameters were measured by TEG, R ranging from 0.9 to 7 minutes, ACT ranging from 2.4 to 8.7 minutes, and MA ranging from 31.2 to 76.6 mm. For the first two parameters, R and ACT, a strong positive correlation between the OVM and TEG was observed ($p < 0.001$), confirming the accuracy of measurement to clotting time. The OVM parameters R and ACT bore a linear relationship with the corresponding TEG values with correlation coefficients, r, of 0.80, 0.73, respectively. In contrast, for the MA parameter, a correlation between the OVM and TEG was not found in the experiment. Many possible factors cause the deviation, one of the most is likely due to differences in the moduli measured for clots formed under quiescent conditions by the OVM, versus conditions of mechanical strain by TEG as discussed below.[16]

As shown in Fig. 6, the Spatio-temporal analysis of MSD reflected variations in the viscoelastic properties of blood during coagulation. The color maps of MSD at minutes 0, 2, 2.5, 3, 7 after activating blood coagulation show that MSD decreased as the coagulation reaction proceeds, which is consistent with the conclusion in Fig. 2.

## 4 Discussion

Laser speckle is a granular intensity pattern produced by the interference of coherent light undergoing multiple scattering from the randomly distributed particles. Laser speckle techniques have been applied to investigate biomedical processes for many years. For example, laser speckle contrast imaging has been widely accepted as an effective method for imaging the blood flow and blood perfusion by measuring the speckle contrast spacetime variations induced by the fast motion of red blood cells in brain superficial blood vessels.[17] In biomechanics evaluation, the range of particle motion is one of the primary parameters, which is governed by the viscoelastic susceptibility of the microenvironment surrounding the particles. LSR has been successfully used to measure the viscoelastic properties of various tissue, such as coronary plaques,[31] human osteoarthritic knee cartilage,[15] and breast tumors,[32] that utilizes the time-varying speckle intensity fluctuations resulting from the random motion of light scattering particles in confined geometries. Therefore, the random optical phase distribution of numerous scattered light waves arising from the motion of scattering particles can be accessed to recover the structured knowledge of the biological tissue.

It is thus possible in principle to characterize the interaction of the wave through the sample by tracking the displacement of specific features of the speckle pattern in time.[33] As the generic and distinctive structures of a speckle pattern, the optical vortex or phase singularities can be used to characterize the viscoelastic properties of the tissue. What's more, the density of the optical vortex is the same order as the speckle spots in a fully developed speckle, and the vortices can determine the structure of the speckle grains since they instruct the sizes and locations of zones of constant phase.[34] Researchers have proposed to exploited first-order statistics of optical vortex motion to measure nanometric displacement and analyze biological kinematic information.[26,27] Like laser speckle rheology and dynamic light scattering investigating the motion of light scatterers, we can explore the second-order stochastic information such as mean square displacement of the optical vortex to obtain the dynamic features of the medium.

In this study, we have proposed and investigated a novel method to evaluate the dynamics process of coagulation by statistics on the motion characteristics of the optical vortex. The result of Fig. 2 showed that the movement pattern of the optical vortex under restrictions during blood coagulation, which is similar to the scattering particles. It maybe opens new opportunities for the field of particles' motion. Moreover, comparative experimental results of the OVM and TEG demonstrated the capability of the OVM to evaluate blood coagulation status. Blood coagulation parameters such as R and ACT measured by the OVM showed a strong correlation with TEG. And the time of global coagulation profiling measured by the OVM is less than that of TEG, as shown in Fig. 3. In addition, the simpler optical devices and low interference with the sample are also important advantages for assessing a patient's blood coagulation status efficiently enough.

A strong correlation between the MSD of the optical vortex and the viscoelastic modulus, |G*| that was measured by a conventional mechanical rheometer for phantoms and tissue samples over a wide range of mechanical moduli have been reported in the previous study. Here, we investigate the accuracy and sensitivity of the method to estimate the coagulation status of porcine blood samples. In un-clotted blood, RBCs and platelets are the most dominant scattering particles, with random and rapid motion, maintaining stable optical properties. When coagulation is initiated, soluble fibrinogen is transformed into insoluble fibrin and polymerizes with activated platelets and RBCs, which leads to scattering particles undergoing restricted Brownian excursions. The

displacement of scattering particles can be defined by the MSD that is related to the viscoelastic modulus of the sample via the GSER as follow:

$$G^*(\omega) = \frac{K_b T}{a\pi \langle \Delta r^2(1/\omega) \rangle \Gamma[1+\alpha(\omega)]} \tag{6}$$

Where $K_b$ is the Boltzmann constant, $T$ represents the temperature in Kelvin. $a$ is the average radius of the light scattering particle. $\langle \Delta r^2(1/\omega) \rangle$ defines the MSD of the light scattering particle at time $1/\omega$. $\Gamma$ denotes the gamma function. $\alpha(\omega)$ is given by

$$\left| \alpha(\omega) = \frac{dln\langle \Delta r^2(t) \rangle}{dlnt} \right|_{t=\frac{1}{\omega}} \tag{7}$$

However, the average radius of the light scattering particle, i.e. $a$, changes continuously with the blood coagulation. The optical methods based on microrheology thereby is difficult to obtain the quantified values of the viscoelastic modulus. Furthermore, the methods require measurement of the scattering coefficient, $\mu_s$, that is evolved continuously as coagulation progressed of the blood sample. To avoid omitting many parameters, the MSD of actual scattering particles is represented by the MSD of the optical vortex. As a virtual particle, it is no sense to measure its average radius. Its MSD can be calculated directly by the method mentioned above and used to characterize the viscoelasticity of the medium, which is verified in the previous study.

To calculate the MSD of the optical vortex accurately, the acquisition frame rate should be selected suitably. Since the rapid change of laser speckle during blood coagulation, the low acquisition frame rate is unable to capture the trajectory of the optical vortex. In this study, the high frame rate of 800 frames per second was chosen for capturing the rapid speckle fluctuations induced by a cascade of coagulation processes. Although higher frame rates are theoretically more useful for capturing the trajectory of the optical vortex, the acquisition frame rate is limited by some hardware parameters such as spatial resolution and exposure time. Small spatial resolution may cause an insufficient number of the optical vortex to be ensemble averaging resulting in unreliable MSD results. Exposure time directly affects the contrast of the image, which is important for the location of the optical vortex. While ensuring sufficient contrast in the image, the exposure time can be reduced by increasing the gain value, thus allowing for a higher frame rate, but this causes more noise and reduces the SNR. A significantly reduced SNR could cause trouble for locating and pairing the optical vortex. Thus, the balance between frame rate and other hardware parameters is essential for the accurate calculation of the optical vortex's MSD.

The captured speckle intensity distribution is represented by real values while the optical vortex is located in the corresponding phase map. It is common practice in physics and engineering that the complex-valued signals are used to represent the real-valued. A true phase map is unable to be obtained by using the simple optical device in this study. We have demonstrated that the MSD of the optical vortex calculated by using pseudo-phase is a strong correlation with it by using real phase in the previous simulation experiments. Therefore, the Laguerre-Gaussian transform was applied to the 2-D speckle pattern, subsequently, calculate the pseudo-phase according to Equation (4) in this study. Several reports have shown that the Riesz transform and Hilbert transform could also be used for obtaining the pseudo-phase, but the Laguerre-Gaussian transform with some irreplaceable advantages was chosen eventually for this study.[21] It can readily be seen from Eq. (3) that the amplitude of the Laguerre-Gaussian function is doughnut-like, which can suppress high-frequency components of unstable phase singularities, thus serves as a band-pass filter. As the bandwidth of this filter, $\omega$ can be used to adjust the density of phase singularities in the pseudo-phase by controlling the average speckle size. Too many optical vortexes may cause the recurrence and collisional disappearance of the optical vortex under the same position, resulting in more pairing errors. In the other case, the insufficient number of the optical vortex can lead to inaccurate statistical results. Thus, it is necessary to choose the suitable value of $\omega$ for the calculation of MSD.

The disappearance of the optical vortex resulting from the collision of positive and negative vortices is a disruption for the statistical calculation of MSD. The random distribution properties of the optical vortex derived from the intensity of laser speckle are the main cause of this problem. During blood coagulation, the rapid change in laser speckle increases the possibility of optical vortex' collisions, which cause the optical vortex to only existing for a very short period. Thus, when the statistical time for the optical vortex's motion increases, the MSD will be unreliable because of the insufficient number of the optical vortex. As shown in Fig 2, we conducted a statistical analysis of all sequences with an acquisition time of 0.5s. The MSD trace that can characterize the viscoelasticity of the blood was obtained at the statistical time, t =0.04s. Moreover, in the initial stage of coagulation reaction(2-3 minutes), the existence time of the optical vortex becomes shorter due to faster fluctuations of laser speckle. When t >0.04s, the fluctuation of the MSD will increase, resulting in a lower correlation with the viscoelasticity of the blood. To investigate a strong correlation with TEG, the MSD data at the moment of 0.02s was chosen eventually to indicate clot viscoelasticity for the whole coagulation process.

The coagulation parameters, R and ACT, that were defined by the OVM presented a strong correlation with corresponding TEG values while the MA parameter demonstrated a weak correlation. It may be likely explained by the differences in the measurement mechanisms of the two methods. The OVM is an optical method that can measure the blood viscoelasticity without mechanical contact during coagulation, causing less mechanical disruption of the clot. In contrast, in TEG, oscillatory shear stress is using for measuring the viscoelasticity during the whole coagulation process, disrupting the clot structure during polymerization and stabilization, thereby causing an obvious impact on MA. Furthermore, the MA parameter is related to stiffness measured for clots formed under quiescent conditions, which is connected with the amount of calcium chloride and blood used. In the OVM, t=0.02s, i.e., a high shearing frequency $\omega$=50 Hz was used for the measurement of different reagent ratios while the low shearing frequency of TEG works also only on the measurement of a fixed ratio reagent. Thus, The absence of a dose-dependent modulation is a possible cause of low MA correlation for both methods. On the other hand, the fluctuations of the clotting curve in Fig. 3. (A) measured by the OVM may show the complex in vivo environment of whole blood coagulation.

In the past, devices such as TEG and ROTEM have been investigated for efficient detection of coagulation in the point of care. In these methods, a cylindrical cup containing a whole blood sample oscillates continuously and a pin on a torsion wire is suspended in the blood. The torsion wire is a sensor that can transmit the changes in blood viscoelastic to an electromagnetic transducer, implementing a coagulation detection function.[9] However, the continuous mechanical oscillation will disrupt the micro-clots structure in the early stages of clotting, enabling them to only evaluate the viscoelastic modulus of larger clots effectively.[11] Thus, the methods described above are less sensitive to tiny and localized changes of blood viscoelasticity and detect the clot formation only after a significant change of blood viscoelastic properties over the entire blood sample. By comparison, in the OVM, the measurements of the optical vortex are related to the nanometer-scale motion of scattering particles. Therefore, the new approach is theoretically sensitive enough to the small changes in clot viscoelasticity and can measure micro clots earlier than TEG and ROTEM. In the study, the R time measured by the OVM ranged from 0.19 to 3.7 minutes while the corresponding value measured by TEG ranged from 0.9 to 7 minutes. Likewise, the ACT measured by the OVM (0.65-7.9 minutes) is less than TEG( 2.4-8.7 minutes). The results demonstrated a shorter clotting time (R and ACT) compared with TEG, which proves the conclusion of the high sensitivity to early incipient micro clots in the OVM.

To overcome the limitations of TEG and ROTEM, other optical approaches for point of care coagulation assessment have been recently reported.[12-14,16] LSR is one of these methods. It characterizes the viscoelasticity of blood by measuring the time scale of speckle intensity fluctuations, $g_2(t)$ to determine blood coagulation status in patients. However, the measurement of $g_2(t)$ in LSR theory conduct little consideration for static scattering, ordered or disordered motion of the scattering particles, the number of times the coherent light is scattered, the offset caused by noise, etc,[17] which is indispensable in practical measurements of $g_2(t)$.

In this study, we validated the capacity of the OVM for the non-contact and rapid measurement of blood coagulation dynamics to assess coagulation status, using just a drop or two of blood. Moreover, the cheap and simple setup that can be easy to operate is also a noteworthy advantage for assessing a patient's blood coagulation status at the bedside. In the future, clinical experiments will be conducted for further correlation studies to validate the potential as a miniaturized medical device for coagulation testing. The OVM may open a powerful opportunity for self-testing and real-time detection of coagulation.